\documentclass[preprint,authoryear,12pt]{elsarticle}

 \usepackage{graphicx}
\usepackage{amssymb}
 \usepackage{lineno}

\journal{Advances in Space Research}

\begin{document}

\begin{frontmatter}

%% Title, authors and addresses

%% use the tnoteref command within \title for footnotes;
%% use the tnotetext command for the associated footnote;
%% use the fnref command within \author or \address for footnotes;
%% use the fntext command for the associated footnote;
%% use the corref command within \author for corresponding author footnotes;
%% use the cortext command for the associated footnote;
%% use the ead command for the email address,
%% and the form \ead[url] for the home page:
%%
%% \title{Title\tnoteref{label1}}
%% \tnotetext[label1]{}
%% \author{Name\corref{cor1}\fnref{label2}}
%% \ead{email address}
%% \ead[url]{home page}
%% \fntext[label2]{}
%% \cortext[cor1]{}
%% \address{Address\fnref{label3}}
%% \fntext[label3]{}

\title{CoRoT and \textit{Kepler} results: solar-like oscillators}

%% use optional labels to link authors explicitly to addresses:
%% \author[label1,label2]{<author name>}
%% \address[label1]{<address>}
%% \address[label2]{<address>}

\author{S. Hekker}

\address{Astronomical Institute `Anton Pannekoek', University of Amsterdam, Science Park 904, 1098 XH Amsterdam, The Netherlands\\
email: S.Hekker@uva.nl, tel: +31 20 525 7491, fax: +31 20 525 7484}

\begin{abstract}
The space-borne observatories CoRoT (Convection Rotation and planetary Transits) and \textit{Kepler} have provided photometric time series data of unprecedented precision for large numbers of stars. These data have revolutionized the fields of transiting exoplanets and asteroseismology. In this review some important asteroseismic results obtained using data from the CoRoT and \textit{Kepler} space missions concerning stars that show solar-like oscillations are discussed. These results comprise, among others, measurements of the location of the base of the convection zone and helium second-ionization zone in main-sequence stars, the presence (or not) of core-helium burning in red-giant stars, as well as differential rotation in these stars.

\end{abstract}

\begin{keyword}
%% keywords here, in the form: keyword \sep keyword

%% MSC codes here, in the form: \MSC code \sep code
%% or \MSC[2008] code \sep code (2000 is the default)
stars -- main-sequence; stars -- subgiants; stars -- red giants; methods -- asteroseismology
\end{keyword}

\end{frontmatter}

 \linenumbers

%% main text

\section{Introduction}
\label{intro}

Internal structure and compositions of stars as well as changes with evolution are broadly understood, but still have many challenges. Classical photometric and spectroscopic observations do not expose the stellar interior due to the extreme opacity of stellar material which masks the internal structure from view. For the Sun the 5-minute acoustic oscillations allow us to peer inside the Sun (helioseismology) and provide much needed information about the internal structure of the Sun and the relevant physical processes. We expect that oscillations in distant stars will also provide invaluable probes of the internal structures of these stars. Oscillations in distant stars have been measured from the ground \citep[e.g.][]{kjeldsen1995b,carrier2003,bedding2004,arentoft2008}. However, these observations have limitations with respect to 1) the length and continuity of the observations, 2) number of stars that can be observed {(mostly relevant for spectroscopy, i.e. high-resolution spectrographs used for asteroseismology can only observe one star at a time, while in photometry many stars can be observed during one integration)}, and 3) the signal-to-noise ratio of the data (mostly relevant for photometry of low-amplitude oscillations). Space missions have distinct advantages that overcome most of these problems. Therefore, dedicated space-missions have been developed. The first asteroseismic space mission MOST \citep[Microvariability \& Oscillations of STars,][]{matthews2000} was launched on June 30, 2003. MOST is a Canadian microsatellite observing bright stars (down to 6th magnitude) with a {photometric} precision of one part per million. In recent years major contributions in asteroseismology have come from the space missions CoRoT and \textit{Kepler}.
The CoRoT\footnote{The CoRoT space mission was developed and operated by the CNES, with participation of the Science Programs of ESA, ESA's RSSD, Austria, Belgium, Brazil, Germany, and Spain.} \citep[Convection Rotation and planetary Transits,][]{baglin2006} and \textit{Kepler}\footnote{\textit{Kepler} is a NASA discovery class mission, whose funding is provided by NASA's Science Mission Directorate.} \citep{borucki2008} space missions have provided unprecedented photometric observations that allowed for many breakthroughs in both stellar and exoplanet research. Some details and characteristics of these missions are outlined here.

\subsection{CoRoT}
The objectives of the CoRoT mission were two-fold: discovering exoplanets down to super-earth radii and unravelling the internal structures of stars.
To achieve this the CoRoT spacecraft was equipped with a 27 cm-diameter afocal telescope with an 8 square-degree field of view. To discover exoplanets via the transit method (when a planet crosses in front of its parent star as viewed by an observer) nearly twelve thousand stars of magnitude 11-16 were observed simultaneously with a cadence of 512 or 32 seconds (i.e., a measurement every 512 or 32 seconds). For seismology about 10 stars with magnitude between 6 and 9 were monitored simultaneously during each run at a cadence of 32 seconds. After the initial observing run of nearly 60 days there were long runs of 150 days alternated with short runs of 20-30 days each monitoring a different patch of the sky {around the so-called CoRoT eyes}.
See \citet{baglin2006}, \citet{auvergne2009} for further technical details of the CoRoT telescope.

CoRoT has observed pre-main-sequence stars \citep[e.g.][]{zwintz2013}, stars on the main-sequence ranging from O and B type stars \citep[e.g.][]{belkacem2009,degroote2010} to stars slightly more massive than the Sun \citep[e.g.][]{michel2008}, as well as more evolved subgiants \citep[e.g.][]{deheuvels2011}, red giants \citep[e.g.][]{deridder2009}, and supergiants \citep{aerts2010sg}. {Additionally, detailed studies of stars harbouring planets have been performed \citep[e.g.][]{ballot2011,gaulme2010}.}

%The CoRoT satellite is restricted to fields  on the sky located near the galactic center and anticenter - often referred to as the `CoRoT eyes'. The CoRoT eyes are 10$^{\circ}$ in diameter and are centered at right ascension = 18h50m and declination = 0$^{\circ}$ for the center field and at right ascension = 6h50m and declination = 0$^{\circ}$ for the anti-center field. The satellite can observe approximately 8 square degrees at any one time.

Launched in December 2006 the CoRoT mission had a nominal lifetime of 2.5 years. It exceeded this nominal period and has taken scientific data till November 2012.

\subsection{Kepler}
The \textit{Kepler} mission was designed primarily to detect earth-size planets in or near the habitable zone (region around a star in which it is theoretically possible for a planet with sufficient atmospheric pressure to maintain liquid water on its surface) and to determine {what percentage} of the billions of stars in our galaxy have such planets. To this end a 0.95-meter Schmidt telescope was designed to be able to observe a 105 square-degree field of view. \textit{Kepler} has been launched in 2009 and monitored about 150\,000 stars {at the same time} in a single field (Cygnus-Lyra region) with magnitudes between 9 and 16  and a cadence of 29.4 minutes (so-called long-cadence mode). Stars with oscillation periods longer than about an hour, such as A and F stars  \citep[e.g.][]{uytterhoeven2011}, and B stars \citep[e.g.][]{balona2011b} on the main sequence, red giants \citep[e.g.][]{bedding2010}, RR Lyrae stars \citep[e.g.][]{kolenberg2010} as well as eclipsing binaries \citep[e.g.][]{prsa2011} could be observed in this way.

Aditionally, the \textit{Kepler} satellite was operated in a short cadence mode in which 512 stars could be observed simultaneously at a cadence of 58.85 seconds \citep{gilliland2010}. This allowed for asteroseismic analyses of Sun-like stars and subgiants \citep[e.g.][]{chaplin2010} which show oscillations with periods of the order of a few minutes. Other stars observed in this observing mode are compact oscillators such as sub-dwarf B stars \citep[e.g.][]{kawaler2010} and white dwarfs \citep[e.g.][]{corsico2012}, as well as Ap stars \citep[e.g.][]{balona2011a}.

The nominal life time of the mission was 3.5 years. {Following the nominal lifetime, the scientific exploitation of \textit{Kepler} was extended but a failure in a reaction wheel put the satellite in safe mode in May 2013. Since then scientific data acquisition has ceased. The \textit{Kepler} team is exploring all possibilities to  solve the problem and put \textit{Kepler} back in scientific operation mode.}\newline
\newline
The data of the CoRoT and \textit{Kepler} satellites have already led to several breakthrough results in our understanding of the internal structures of stars as well as in the vast numbers and diversity of detected planetary systems. In the next section a more detailed description of asteroseismology is presented followed by a selection of important asteroseismic results obtained with CoRoT and \textit{Kepler} for low-mass main-sequence stars, subgiants, and red-giant stars. This review finishes with some future perspectives.

\section{Asteroseismology}
\label{astero}
\subsection{Introduction}
Asteroseismology is the study of the internal structures of stars by means of their intrinsic global oscillations. The characteristics of these oscillations are determined by the stellar interior and hence it is possible to infer the stellar structure from the oscillations. Many (but, as far as we know, not all) stars across the Hertzsprung-Russell diagram oscillate. % (see Figure~\ref{asteroHRD}). 

\begin{figure}
\centering
\begin{minipage}{0.2\linewidth}
\centering
\includegraphics[width=\linewidth]{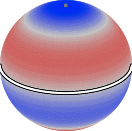}
\end{minipage}
\hfill
\begin{minipage}{0.2\linewidth}
\centering
\includegraphics[width=\linewidth]{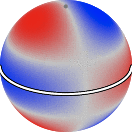}
\end{minipage}
\hfill
\begin{minipage}{0.2\linewidth}
\centering
\includegraphics[width=\linewidth]{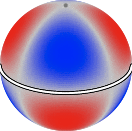}
\end{minipage}
\hfill
\begin{minipage}{0.2\linewidth}
\centering
\includegraphics[width=\linewidth]{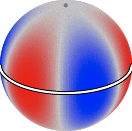}
\end{minipage}
\caption{Schematic representation of  $\ell=3$ oscillation modes with $m=0,1,2,3$ from left to right. The blue areas indicate regions where the star expands while the red parts contract. The gray lines are the nodal lines with the dark gray dot at the top indicating the rotation axis, which lies in the plane of the paper. The equator is indicated in white. Figure obtained from non radial pulsator (NRP) animation creator \citep{schrijvers1997}.}
\label{modeid}
\end{figure}

Stellar oscillations can be characterised by their frequencies, typical lifetime (coherence time), amplitudes and mode identification. {Oscillation modes are identified by} the radial order ($n$) indicating the number of nodes in the radial direction, the spherical degree ($\ell$) indicating the number of nodal lines on the surface, and the azimuthal order ($m$) indicating the number of nodal lines that pass through the rotation axis. Examples of $\ell=3$ modi with different $m$ values are shown in Figure~\ref{modeid}.
The mode identification in terms of $n, \ell, m$, the frequency and amplitude of the modes as well as effective temperature, surface gravity, and metallicity, allow for a study of the internal structure of the star. {The internal stellar structure can be studied by comparing the observational results with results from stellar models, or directly from the observational data as is the case for glitches (see Section 2.3.4). An example of model dependent results is presented by \citet{metcalfe2012}, while \citet{mazumdar2012K} show model independent results.}

There exist mainly two types of oscillators {depending on the} driving mechanism and their resulting mode pattern.

\subsection{Driving mechanisms}
\subsubsection{Heat engine}
Oscillations can be driven by a so-called heat engine. In this mechanism a particular layer in the star gains heat during the compression part of the oscillation cycle. This usually occurs in partially ionized layers of elements with high opacity, such as hydrogen, helium, or iron. Inside these partially ionized layers the energy, which would normally heat the zone, mostly goes into increasing the ionization as the star is compressed. This increases the opacity of the zone, trapping radiation more efficiently, and resulting in outward pressure. The zone is then driven outwards, cooling as it rises, which increases the opacity and outward pressure. Further cooling of the zone results in recombination of the ionized material, and with that a sudden decrease in the opacity decreasing the outward pressure. Hence, the zone drops back and the cycle can start again. {Due to the role of opacity this is also called $\kappa$ mechanism.}

The location of the layer(s) that act(s)Ä as a heat engine is crucially important to effectively drive global oscillations as all other layers that lose heat on compression damp the oscillations. See for more details \citet{aerts2010}.

Oscillators driven by the heat engine, such as Cepheids, have coherent oscillation modes with very long lifetimes and hence their oscillation peaks in the Fourier spectrum resemble $\delta$-functions. The observed oscillations are affected by selection effects and the mode identification has to be obtained from multi-colour photometry (amplitude ratios and / or phase differences), or from high-resolution spectroscopy (line-profile variations).

\subsubsection{Stochastic driving}
Oscillations can be stochastically excited by the {turbulent convection of stars \citep[e.g.][]{goldreich1977,goldreich1988}}. Effectively, some of the {convective energy} is transferred into energy of global oscillations. The oscillations in the Sun are stochastically driven and damped. Hence these type of oscillations are referred to as solar-like oscillations.  It is expected that in all stars with turbulent outer layers such oscillations are excited. For a recent extensive overview of solar-like oscillators see \citet{chaplin2013}. {In addition to the oscillations the convection also produces a convective background on which the p-mode spectrum is seated. See \citet{mathur2011} for a description and characterization of the convective background.}

%The restoring force for stars perturbed from equilibrium can be pressure (p), or buoyancy (g). These lead to p and g-mode oscillations, respectively.

\begin{figure}
\centering
\begin{minipage}{0.8\linewidth}
\centering
\includegraphics[width=\linewidth]{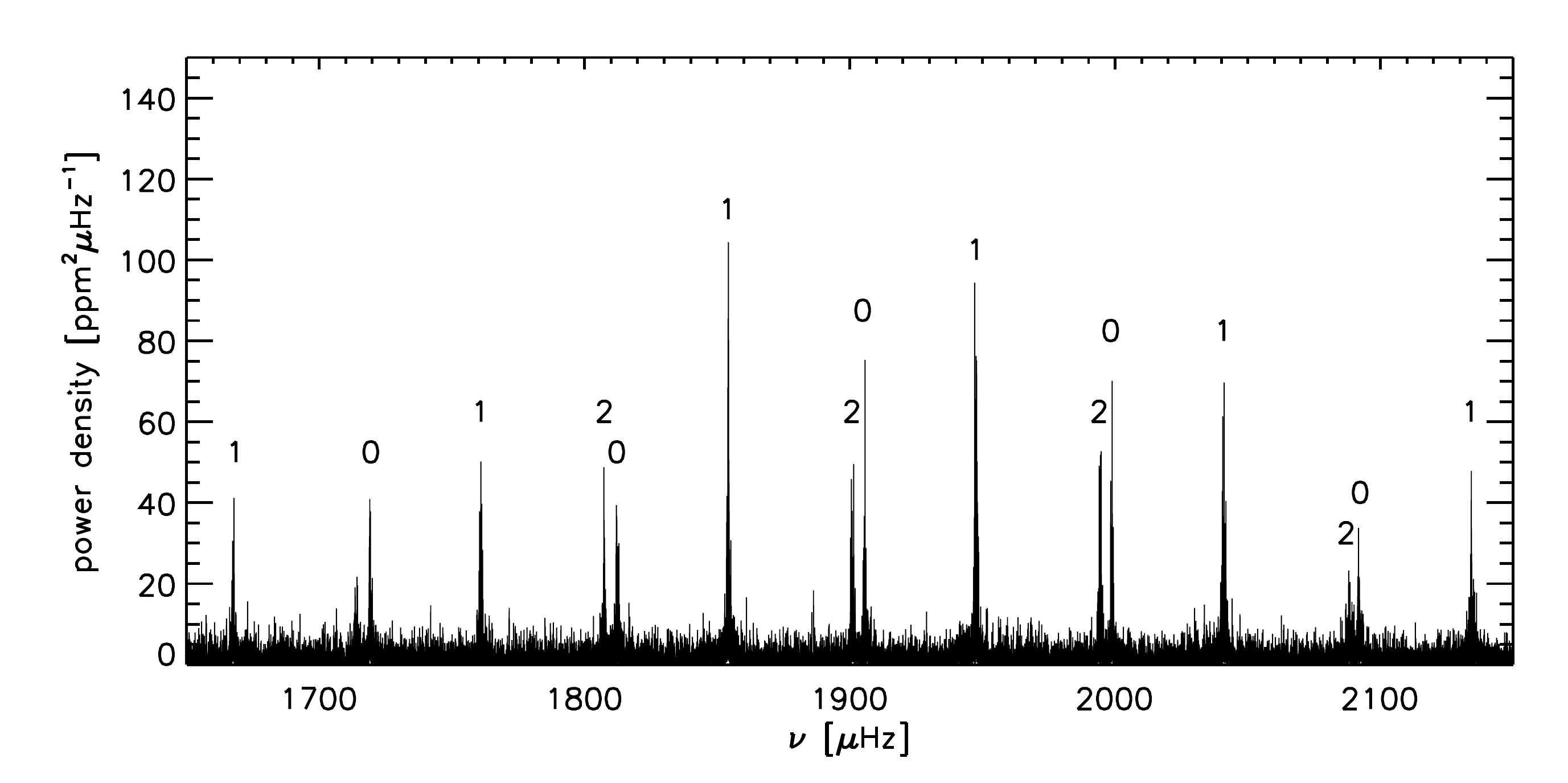}
\end{minipage}
\begin{minipage}{0.8\linewidth}
\centering
\includegraphics[width=\linewidth]{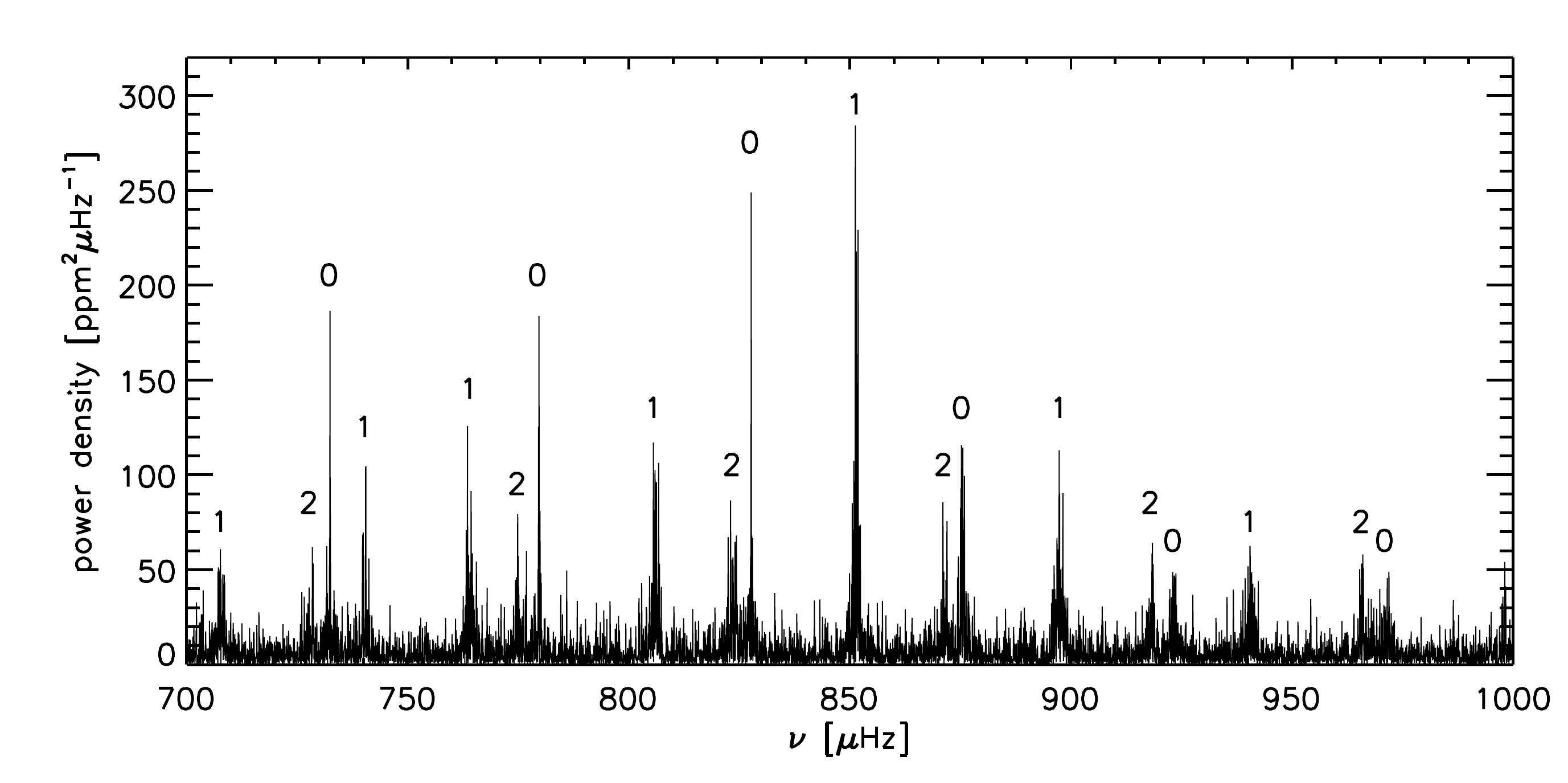}
\end{minipage}
\begin{minipage}{0.8\linewidth}
\centering
\includegraphics[width=\linewidth]{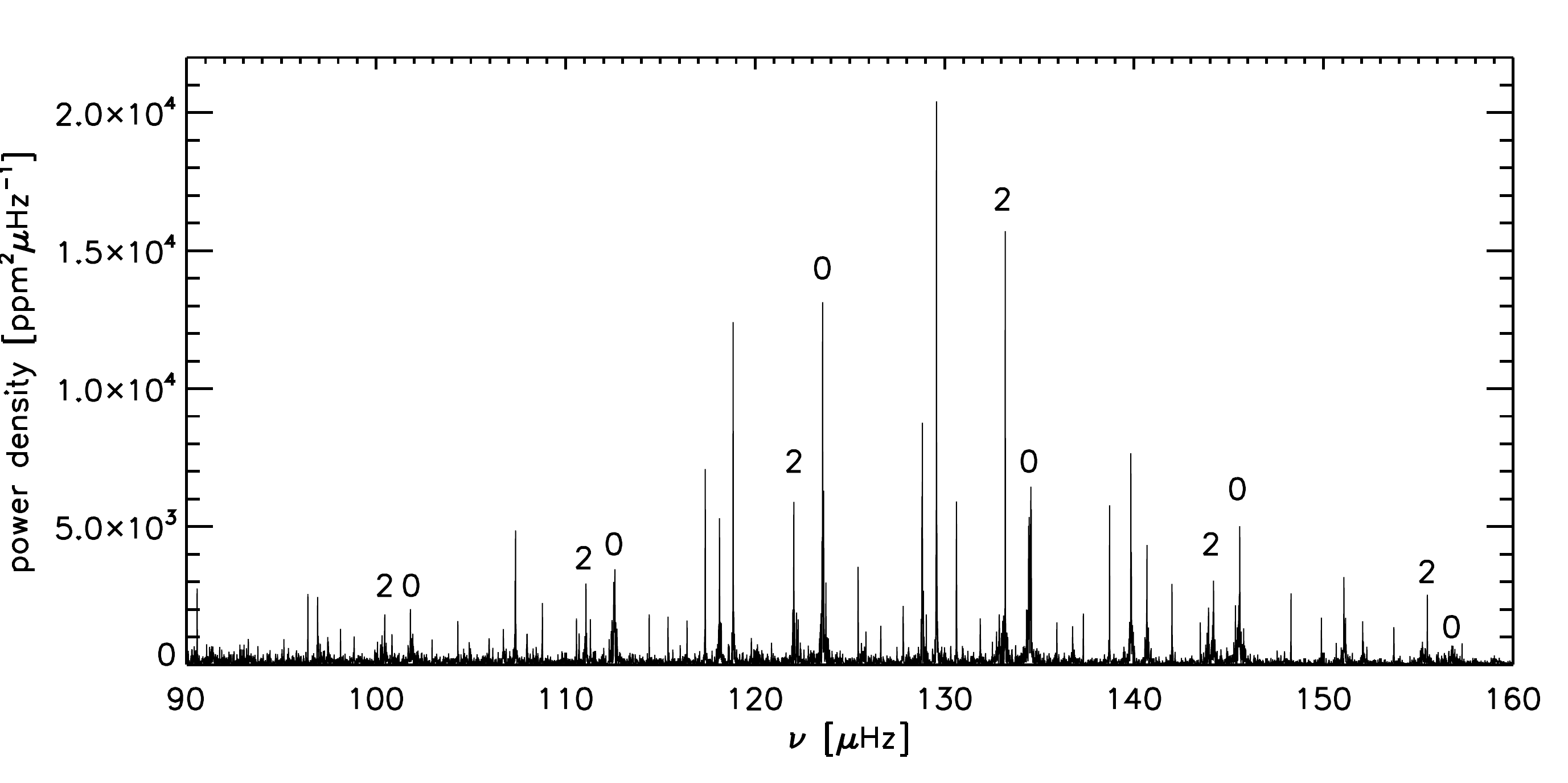}
\end{minipage}
\caption{Fourier power density spectra of a main-sequence star \citep[KIC 3656479 (top),][]{mathur2012}, a subgiant \citep[KIC 11395018 (middle),][]{mathur2011FF}, and a red giant (KIC 9145955, bottom). The numbers indicate the degree ($\ell$) of the modes. The subgiant in the middle panel has a mixed mode pair at 740.3 and 764.0 $\mu$Hz. {This can be identified by the fact that two $\ell=1$ modes are present between the $\ell=0,2$ pairs at $\sim$ 730 and 770 $\mu$Hz.} {The vertical dotted lines indicate $\nu_{\rm max}$ (Eq.~\ref{numax}) and the horizontal dashed-dotted line (top panel) indicates $\Delta\nu$ (Eq.~\ref{dnu}).}}
\label{powerspectra}
\end{figure}

The stochastic driving and damping of solar-like oscillations results in a finite mode lifetime (few minutes to about hundred days) of the oscillation modes. This finite mode lifetime is visible as a typical width in the peaks representing the oscillations in the Fourier spectrum, with shorter lifetimes resulting in wider peaks.  Additionally, in stochastic oscillators essentially all modes are excited albeit with different amplitudes. This results in a clear pattern of overtones in the Fourier spectrum (see top panel of Figure~\ref{powerspectra}). This is of great value for identifying the oscillation modes. \newline
\newline
For the remainder of this review only stars with stochastically-driven solar-like oscillations will be considered.
%These oscillations have however very low-amplitudes which made them challenging to observe in the pre-space photometry and pre-high-precision spectra era. 

%\subsubsection{Characterisation}

\subsection{Solar-like oscillators}
%\subsubsection{Pressure modes}
{Stars with a convective zone that excite stochastic oscillations extend from low-mass ($M\lesssim2$M$_{\odot}$) main-sequence stars till evolved red giants. Stars on the main sequence are those that fuse hydrogen in their core. Then when the core hydrogen is exhausted, the stars start to burn hydrogen in a shell around an inert helium core, while the surface is cooling (subgiants). On the red-giant branch the luminosity and radii of the stars increase up to the onset of helium-core burning. Then the radii and surface temperatures decrease and the helium-core burning stars settle on the horizontal branch (also called red clump).}

%The outer layers of low-mass ($M\lesssim2$M$_{\odot}$)  main-sequence stars (stars fusing hydrogen in their core), subgiants (stars that have exhausted hydrogen in their core and are starting to burn hydrogen in a shell around an inert helium core, while the surface is cooling), and red giants (hydrogen-shell-burning stars ascending the red-giant branch increasing their luminosity and radius as well as stars in their helium-core-burning phase on the horizontal branch (also called red clump)) are turbulent and these stars possess solar-like oscillations. 

Oscillations can be trapped in a cavity in the outer layers of the star where the restoring force is pressure and the oscillations have the nature of standing acoustic waves. These oscillation modes are referred to as p-mode oscillations. The oscillations can also be trapped in an inner cavity in the inner radiative region of the stars where the restoring force is buoyancy and the oscillations have the character of standing gravity waves. These are called g-mode oscillations. 

%In the Sun the cavity of low degree p modes reaches maximum depths of about 0.3 solar radius, while the g modes are trapped in the core below 0.2 solar radius. 
{In the Sun the lower turning points of low-degree non-radial modes --that can be observed in other stars-- are roughly between 0.05 and 0.2 solar radius, depending on the frequency and degree of the mode \citep[e.g.][]{lopes1994,garcia2008}. Note, however, that the non-radial p modes are most sensitive to the part of the cavity where they spend most time, i.e., the part where the sound speed is lowest. The sound speed decreases towards the surface, and hence the oscillations carry most information of the outer part of the cavity.
The g modes are most sensitive in a radiative region deeper in the star below the lower turning point of the p modes. The dipole asymptotic g-mode spacing in the Sun was measured by \citet{garcia2007}. Unfortunately, due to the small amplitude of the g modes at the surface of the Sun, it has not been possible to detect g modes directly.}
%Individual g modes have not been detected in the Sun because of their small surface amplitudes \citep{garcia2007}. 
When stars evolve to the subgiant phase the frequencies of the oscillations in the p- and g-mode cavities nearly overlap and a few so-called mixed modes can be observed (see middle panel of Figure 2). Such modes have g-mode nature in the {radiative interior} and p-mode nature in the outer cavity. For further evolved red-giant stars a large number of frequencies in the g-mode cavity overlap with the oscillation frequencies in the p-mode cavity and many mixed modes can be observed (see bottom panel of Figure~\ref{powerspectra}). 

\subsubsection{Pressure modes}
P-mode oscillations in the Sun have a typical period of about 5 minutes which corresponds to frequencies of roughly 3 mHz. Lower-mass stars on the main-sequence are smaller and have oscillations with shorter periods, i.e., higher frequencies. More evolved stars with larger radii show oscillations with longer periods of hours to days, resulting in frequencies in the sub-mHz or $\mu$Hz regime. The typical {frequency region in which the acoustic modes oscillate can be} characterised by $\nu_{\rm max}$, the frequency of maximum oscillation power {(vertical dotted lines in Fig.~\ref{powerspectra})}. This is an important diagnostic as it scales with the mass ($M$), radius ($R$), and effective temperature ($T_{\rm eff}$) of the star \citep[e.g.][]{brown1991,kjeldsen1995,belkacem2011}:
\begin{equation}
\nu_{\rm max} \propto \frac{M}{R^2 \sqrt{T_{\rm eff}}} \propto \frac{g}{\sqrt{T_{\rm eff}}},
\label{numax}
\end{equation}
with $g$ the surface gravity.

The pattern of the frequencies in the Fourier spectrum can be described asymptotically by \citep{tassoul1980,gough1986}:
\begin{equation}
\nu_{n,\ell} \simeq \Delta\nu(n+\frac{\ell}{2}+\epsilon)-\delta\nu_{n,\ell},
\label{tassoul}
\end{equation}
{where $\nu$ is frequency and $\epsilon$ is a phase term}. $\Delta\nu$ is the regular spacing in frequency of oscillations with the same degree $\ell$ and consecutive orders $n$ {(horizontal dashed-dotted line in the top panel of Fig.~\ref{powerspectra})}. This is proportional to the inverse of the acoustic diameter, i.e. the sound travel time across a stellar diameter. Furthermore, it can be shown that $\Delta\nu$ is proportional to the square root of the mean density ($\overline{\rho}$) of the star \citep{ulrich1986}:
\begin{equation}
\Delta\nu = \nu_{n,\ell} - \nu_{n-1,\ell} = \left(2 \int \frac{{\rm d}r}{c} \right )^{-1} \propto \sqrt{ \overline{\rho}},
\label{dnu}
\end{equation}
with $c$ the adiabatic sound speed and $r$ the distance to the centre of the star. {Note that slightly different methods to measure $\nu_{\rm max}$ and $\Delta\nu$ are available \citep[e.g.][]{huber2009,mosser2009,hekker2010method,kallinger2010,mathur2010}. It has been shown that the different determinations are consistent within their uncertainties and definitions \citep[e.g.][]{verner2011,hekker2011comp,hekker2012}.}

Finally, $\delta\nu_{n,\ell}$ is the so-called small frequency separation which can be approximated asymptotically by \citep[e.g.][]{gough1986}:
\begin{equation}
\delta\nu_{n,\ell} = \nu_{n,\ell}-\nu_{n-1,\ell+2} \simeq -(4\ell+6)\frac{\Delta\nu}{4\pi^2 \nu_{n,\ell}}\int_0^R\frac{{\rm d} c}{{\rm d} r}\frac{{\rm d} r}{r}.
\label{smallsep}
\end{equation}
For main-sequence stars Eq.~\ref{smallsep} is rather strongly weighted towards the central parts of the star. Hence $\delta\nu_{n,\ell}$ provides a measure of the helium content in the core. In the core of more evolved stars the increase in the helium content causes a reduction in the sound speed, and hence a reduced weight of the central parts of the star in the integral of Eq.~\ref{smallsep}. Therefore in more evolved stars $\delta\nu_{n,\ell}$ does not provide a measure of the helium content in the core.

The near-surface regions of stars are difficult to model properly due to convection. Additionally the oscillations are strongly non-adiabatic in the superficial layers. These near-surface problems cause a shift in the frequencies of stellar models compared to the observed frequencies, which should be accounted for. The large and small frequency separation (Eqs.~\ref{dnu} and \ref{smallsep}) are also affected by the near-surface problems although less severely. To eliminate the influence of the near-surface region, frequency ratios such as $r_{02}$ (Eq.~\ref{fratio}) can be constructed, which are essentially independent of the near-surface problems \citep[e.g.][]{roxburgh2003,roxburgh2005}.
\begin{equation}
r_{02}=\frac{\nu_{n,0}-\nu_{n-1,2}}{\nu_{n,1}-\nu_{n-1,1}}
\label{fratio}
\end{equation}

\subsubsection{Gravity modes}
The g modes in the {inner radiative region} have a high number of radial nodes ($n_{\rm g}$) due to the high density of the core. These oscillation modes follow an asymptotic approximation equivalent to the p modes, but with a regular spacing in period ($\Delta\Pi$):

\begin{equation}
\Delta\Pi_{\ell}=\frac{2\pi^2}{\sqrt{\ell(\ell+1)}} \left( \int\frac{N}{r} dr \right) ^{-1}
\label{dp}
\end{equation}
with $N$ the so-called buoyancy frequency, i.e., the characteristic frequency for internal gravity waves \citep[e.g.][]{tassoul1980}.

\subsubsection{Mixed modes}
In subgiants and red giants the frequencies in both the p-mode cavity in the outer atmosphere and the g-mode cavity in the inner radiative region have similar values and a coupling between the cavities can persist. In that case a p and g mode with similar frequencies and same spherical degree undergo a so-called avoided crossing. The interactions between the modes will affect (or bump) the frequencies. This bumping can be described as a resonance interaction of two oscillators \citep[e.g.][]{aizenman1977,deheuvels2010,benomar2013}. This causes additional non-radial ($\ell > 0$) mixed modes with dual p-g character to appear in the near regular p-mode frequency pattern described above (see middle and bottom panel of Figure~\ref{powerspectra}). These mixed modes show a predictable pattern in period spacing originating from the g-mode nature and the bumping with the p modes.

\subsubsection{Sharp changes in stellar properties: Glitches}
Information on specific transition regions in a star, such as the boundaries of convective zones or ionization zones of helium or hydrogen, can be obtained from the fact that, at such boundaries, the properties of the star change on a scale substantially smaller than the local wavelength of the oscillations \citep[e.g.][]{vorontsov1988,gough1990}. These sharp features (also called glitches) cause oscillatory variations in the frequencies with respect to the pattern described in Eq.~\ref{tassoul}, where the period of the variation depends on the location of the feature, while its amplitude depends on the detailed properties of the feature. Note that the determination of glitches is completely independent of stellar models.

\subsubsection{Rotation}
Stellar rotation lifts the degeneracy of non-radial modes, producing a multiplet of $(2\ell+1)$ frequency peaks in the Fourier spectrum for each mode. The frequency separation between two mode components of a multiplet is related to the angular velocity of the stellar material in the propagation region, {under the assumption of the absence of magnetic splitting}. This allows us to derive the rotation rate of the star in the region to which the mode is most sensitive. The relative height of the mode components in a multiplet is sensitive to the inclination angle at which the star is viewed \citep{gizon2003}.\newline
	
Note that there are several excellent review papers by Christensen-Dalsgaard \citep[e.g.][]{jcd2004,jcd2011,jcd2012} which provide a thorough theoretical overview of solar-like oscillations.

\section{Recent results from CoRoT and \textit{Kepler}}
In this section a selection of the results obtained with CoRoT and \textit{Kepler} for main-sequence stars and red-giant stars is presented.
\subsection{Main-sequence stars}
\subsubsection{F-stars}
Early results from CoRoT came from a number of preselected main-sequence stars that are slightly hotter than the Sun \citep{michel2008} of which HD49933 became the best known `prototype'. This star is an F5V star about 1000 K hotter than the Sun. In these hotter stars the stochastic oscillations have shorter lifetimes, which results in wide peaks in the Fourier spectrum. Due to this effect the $\ell=2$ and $\ell=0$ modes might overlap in their frequency values. This causes difficulties for the mode identification \citep[e.g.][]{appourchaux2008,benomar2010}. This phenomenon is also seen in other F stars observed with CoRoT \citep[e.g.][]{barban2009,garcia2009} and \textit{Kepler} \citep[e.g.][]{appourchaux2012a,appourchaux2012b} as well as for oscillations in F-stars observed with radial velocity measurements such as Procyon \citep{hekker2008,arentoft2008,bedding2010,huber2011procyon}. \citet{white2012} have now devised a method based on the phase term $\epsilon$ (see Eq.~\ref{tassoul}), which seems to be able to identify the correct scenario.

%The first data from CoRoT for HD49933 were analysed using well tested Maximum Likelihood Estimation methods \citep{appourchaux2008}. This analysis was then revised using posteriori probabilities \citep{appourchaux2009} and Bayesian techniques \citep{gaulme2009,benomar2009a}. To find a definitive answer for the mode identification, HD49933 was observed during an additional long observing run. This enabled firm mode identification \citep{benomar2009b} as well as the detection of a stellar activity cycle resembling the solar 11-year cycle with its impact on the p-mode oscillations \citep{garcia2010,salabert2011} and sharp internal structure features \citep{mazumdar2012}.

%The mode identifications of F stars have been debated for all stars in this class. Distinguishing between two identified scenarios has been difficult even when using detailed modelling. The models would always favour one scenario, but often not by much. 

\subsubsection{Planet-hosting stars}
The power of asteroseismology for planet-hosting stars lies in the determination of accurate stellar parameters. From the detection of exoplanets using the transit method as done by CoRoT and \textit{Kepler} it is possible to obtain the planetary properties as a function of the host-star properties. Hence, the accuracy of, for instance, the determined planetary radius depends crucially on the stellar radius. Stellar radii from asteroseismology have typical uncertainties of the order of 2-3\%  and are thus far better than can be obtained from the position in the H-R diagram, which is often the alternative. Such accurate stellar properties have led to breakthrough results, {such as CoRoT-7b, the first transiting super-earth \citep[e.g.][and references therein]{leger2009,wagner2012}; Kepler-21b, a 1.6 earth-radius planet \citep{howell2012}}; Kepler-36, a system with a pair of planets with neighbouring orbits and dissimilar densities \citep{carter2012}; or Kepler-37b, a sub-Mercury sized planet \citep{barclay2013}.

From asteroseismology it is also possible to obtain the stellar rotation and the inclination angle, i.e., the angle between the rotation axis and the observer \citep[e.g.][]{ballot2011}. These parameters can be used to study the obliquity of exoplanet host stars, i.e., the angle between the stellar spin axis and the planetary orbital axis. \citet{chaplin2013obl} present such a study for two planetary systems and find that coplanar orbits are favored in both systems, and that the orientations of the planetary orbits and the stellar rotation axis are correlated.
{\citet{huber2013} revisited all \textit{Kepler} planet-host stars showing pulsations and refined the fundamental properties of these planet-candidate host stars.}

\subsection{Evolved stars}

\begin{figure}
\begin{minipage}{\linewidth}
\centering
\includegraphics[width=\linewidth]{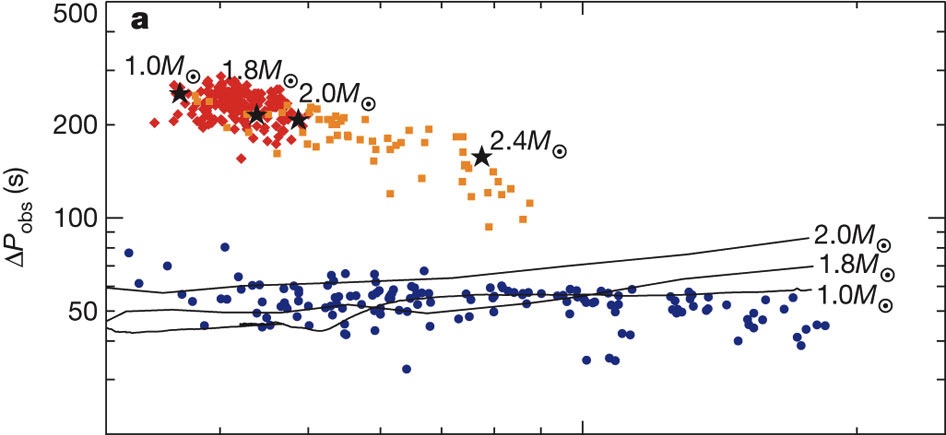}
\end{minipage}
\begin{minipage}{\linewidth}
\centering
\includegraphics[width=\linewidth]{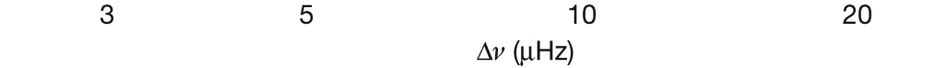}
\end{minipage}
\caption{Observed period spacing ($\Delta P$) as a function of large frequency separation ($\Delta\nu$) with stars in the hydrogen-shell-burning phase on the red-giant branch shown in blue dots and the stars in the helium-core-burning phase shown in red diamonds and orange squares. The solid lines show average observable period spacings for ASTEC \citep{jcd2008} models of hydrogen-shell-burning giants on the red-giant branch as they evolve from right to left. The black stars show theoretical period spacings calculated in the same way, for four models of helium-core-burning stars that are midway through that phase (core-helium fraction 50\%). The 2.4~M$_{\odot}$ model was calculated with ASTEC and commenced helium-burning without passing through a helium flash. The other three models, which did undergo a helium flash, were computed using the ATON code \citep{ventura2008}. Solar metallicity was adopted for all models, which were computed without mass loss. Figure adapted from \citet{bedding2011}.}
\label{dpfig}
\end{figure}

\subsubsection{Non-radial oscillations}
Before the first results of CoRoT it was debated whether non-radial oscillations would be observable for red-giant stars. It was thought that non-radial oscillations would be damped in the inner g-mode cavity. This was based on the analysis of WIRE data of the K0III star $\alpha$ UMa \citep{buzasi2000} and the subsequent theoretical interpretation \citep{dziembowski2001}. Observations in a radial velocity multi-site campaign of two other stars, $\xi$ Hydrae and $\epsilon$ Ophiuchi, were interpreted accordingly \citep{frandsen2002,stello2006,deridder2006,mazumdar2009}.
However, investigation of spectral-line shapes of these stars revealed that non-radial oscillations are detectable in red-giant stars \citep{hekker2006}. 
Data from the CoRoT mission confirmed this unambiguously \citep{deridder2009} with hundreds of stars showing a Fourier spectrum with a similar pattern of $\ell=0,1,2$ modes as present in the Sun. At the same time this was also understood from theoretical calculations by \citet{dupret2009}. Not only did these authors predict that non-radial oscillations with predominantly p-mode nature could reach observable heights in the Fourier spectrum, they also predicted the presence of additional g-dominated mixed modes. %\citet{dupret2009} showed that modes with large amplitudes in the {inner} g-mode cavity and a frequency close to a `pure' p mode of the same degree can reach observable heights at the surface of the star. 

Gravity-dominated mixed modes in red-giant stars have recently been discovered observationally for $\ell = 1$ modes by \citet{beck2011}. From these mixed modes the period spacings (see Eq.~\ref{dp}) can be derived. \citet{bedding2011} and \citet{mosser2011mm} showed that the observed period spacings of red giants in the hydrogen-shell-burning phase ascending the red-giant branch is significantly different from the period spacing for red giants also burning helium in the core. This provides a clear separation into these two groups of stars that are superficially very similar (see Figure~\ref{dpfig}). This effect was explained by \citet{jcd2011} as being due to convection in the central regions of the core in the helium-burning stars. The buoyancy frequency ($N$) is nearly zero in the convective region. As the period spacing is inversely proportional to the integral over $N$ (see Eq.~\ref{dp}), this increases the period spacing compared to a star without a convective region in the core.

\subsubsection{Rotation}
\begin{figure}
\begin{minipage}{\linewidth}
\centering
\includegraphics[width=\linewidth]{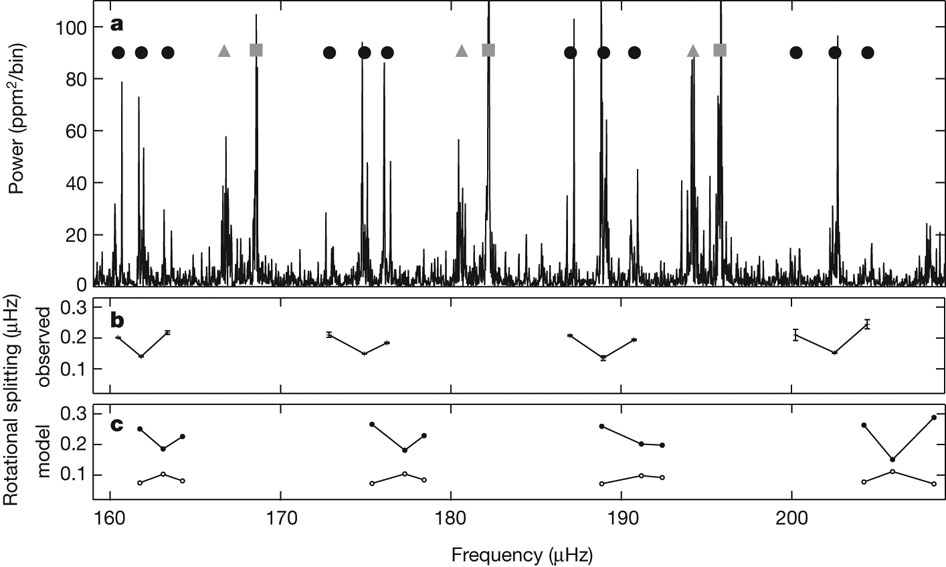}
\end{minipage}
\caption{Top: Fourier spectrum of a rotating red giant. Grey squares indicate radial modes, grey triangles indicate $\ell = 2$ modes, and black circles indicate $\ell = 1$ rotational multiplets. Middle: the observed rotational splitting for individual $\ell = 1$ modes. Bottom: theoretically predicted rotational splitting assuming two different rotation laws. Dots: splitting for non-rigid rotation for the case of a core rotation ten times faster than the surface rotation, which resembles the observations qualitatively well. Open circles: theoretical splittings for rigid rotation, which show a trend opposite to the observed one, with the largest splitting in the centre of the dipole forest and lower splitting in gravity-dominated modes. Figure taken from \citet{beck2012}.}
\label{rotfig}
\end{figure}

Mixed modes are most sensitive in either the {internal} or the outer region of the star depending on their predominant g- or p-mode nature. Hence the characteristics of these modes with different nature can reveal information on properties at different depths in the star. One of these characteristics is the splitting of the mode in $m=-\ell,..,\ell$ azimuthal orders due to rotation. The rotational splitting of dipole {mixed} modes with different p-g nature have been measured \citep{beck2012} and show that for a red-giant branch star the core rotates approximately ten times faster than the surface (see Figure~\ref{rotfig}). This discovery was followed by a result for a low-luminosity red-giant branch star for which the core rotation is about five times faster than the surface rotation \citep{deheuvels2012}.
Subsequently, \citet{mosser2012rot} showed that stars ascending the red-giant branch experience a small increase of the core rotation followed by a significant slow-down in the later stages of the red-giant branch resulting in slower rotating cores in red-clump (or horizontal-branch) stars compared to stars on the red-giant branch. 

\subsection{Glitches}
Only with the data from CoRoT and \textit{Kepler} it has become possible to measure glitches, i.e., the signature of sharp internal structure changes, in other stars than the Sun. \citet{miglio2010} studied the internal structure of the CoRoT target HR 7349 from frequencies obtained by \citet{carrier2010}. \citet{miglio2010} interpreted a periodic component in the oscillation frequencies as caused by a local depression of the sound speed that occurs in the helium second-ionization region at a relative acoustic radius of $\sim$0.55 (in the Sun the helium second-ionization region lies at a relative acoustic radius of $\sim$0.83 and the base of the convection zone at a relative acoustic radius of $\sim$0.4). Subsequently, \citet{mazumdar2012} studied the sharp internal structure changes of HD49933 and were able to find both the signatures of the helium second-ionization zone and of the base of the convective envelope. Similar studies of 19 stars observed with \textit{Kepler} found that the extracted locations of the layers of sharp variation in sound speed are consistent with representative models of these stars \citep[e.g.][]{mazumdar2012K}.

\subsection{Ensemble asteroseismology}

\begin{figure}
\begin{minipage}{\linewidth}
\centering
\includegraphics[width=\linewidth]{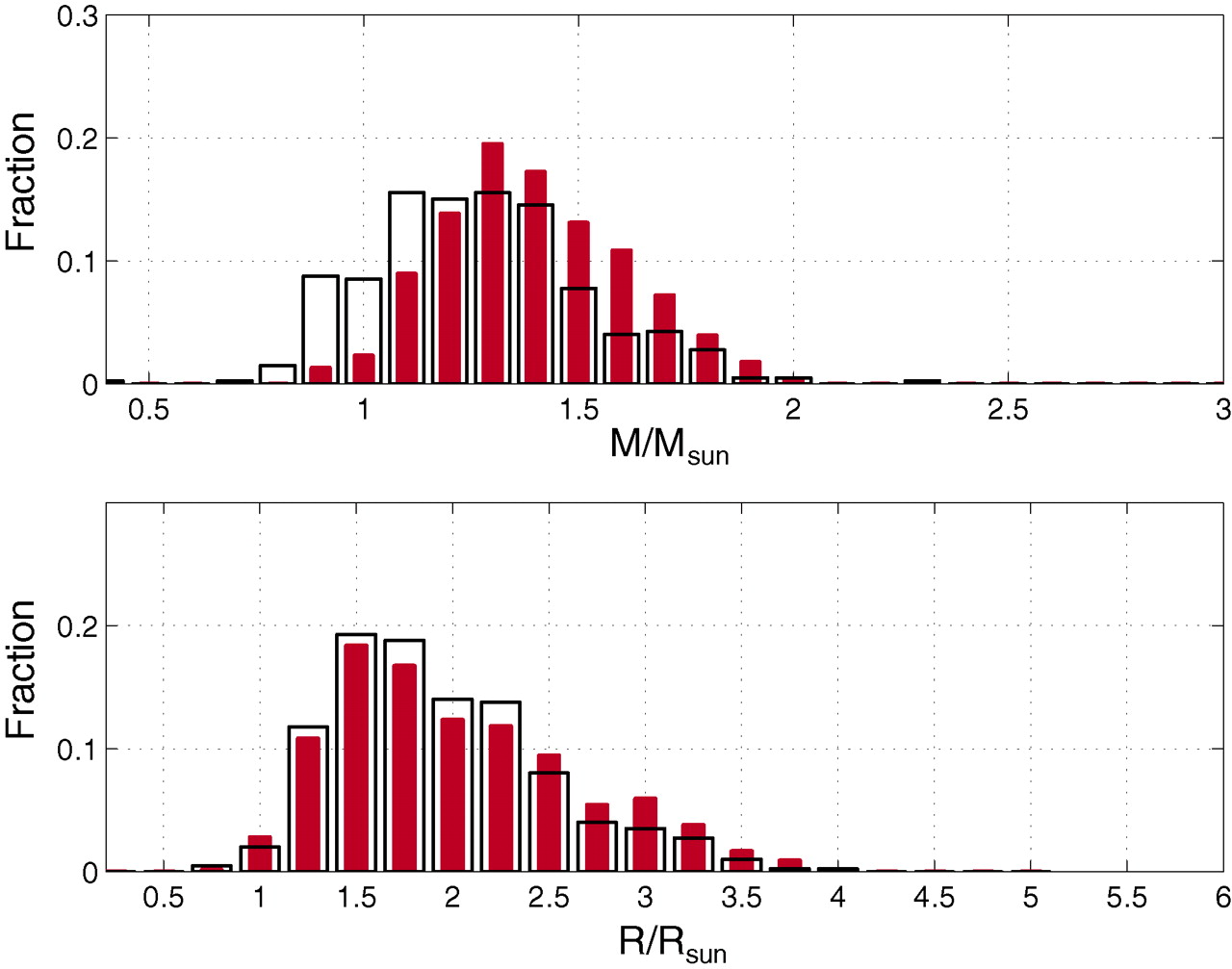}
\end{minipage}
\caption{Black lines show histograms of the observed distribution of masses (top) and radii (bottom) of the \textit{Kepler} main-sequence and subgiants ensemble \citep{chaplin2011}. In red, the predicted distributions from population-synthesis modeling, after correction for the effects of detection bias. The population modeling was performed by using the TRILEGAL code \citep{girardi2005,miglio2009}. Figure taken from \citet{chaplin2011}.}
\label{msdistr}
\end{figure}

Before the CoRoT and \textit{Kepler} era solar-like oscillations were confirmed in 10-20 stars. This number increased significantly with CoRoT and \textit{Kepler} data. For example, with CoRoT the number of red giants with detected solar-like oscillations increased from three to hundreds. These large numbers allowed for the first time to investigate an ensemble of stars as a whole, i.e., to perform statistical studies of intrinsic stellar properties (such as mass and radius) and to test theories of stellar evolution. Such ``ensemble asteroseismology'' studies were performed by e.g. \citet{hekker2009,huber2010,kallinger2010,mosser2010}. These studies led, among others, to observational evidence for a tight correlation between $\nu_{\rm max}$ and $\Delta\nu$ \citep{hekker2009,stello2009,mosser2010}. Additionally, it was found that oscillations in red giants follow a universal pattern \citep{mosser2011}. Furthermore, investigations of properties, such as the amplitudes of the oscillations, as a function of evolution have been presented \citep{huber2010}.  \citet{miglio2009} followed this up with a comparison of the observed distribution of the large frequency separation with distributions from different synthetic populations showing that it is not likely that a recent starburst has taken place in our part of the Milky Way.

With \textit{Kepler} the number of stars with detected solar-like oscillations has been increased to tens of thousands, among which about 500 are main-sequence and subgiant stars. An ensemble study of these main-sequence and subgiant stars was performed by \citet{chaplin2011}. One of the main conclusions resulting from this work concerned the similarity in the observed and simulated radius distributions as well as the difference between the stellar-mass distribution derived observationally from the asteroseismic measurements and the mass distribution of synthetic populations (see also Figure~\ref{msdistr}). On the assumption that the observed masses and radii are robust, this result may have implications for both the currently excepted star-formation rate as well as initial-mass function of stars. Mixing or overshooting of material between different layers (including stellar cores) and the choice of the so-called mixing-length parameter, which measures the typical length scale of the convection, may also be relevant \citep{chaplin2011}.

A more detailed review of ensemble asteroseismology based on CoRoT and \textit{Kepler} data has been presented by \citet{chaplin2013}.

\section{Future}
CoRoT and \textit{Kepler} have both exceeded their minimal lifetime, but are currently out of operation since November 2$^{\rm nd}$ 2012 (CoRot) and mid May 2013 (\textit{Kepler} in safe mode).

Notwithstanding the technical problems at these stages, both CoRoT and \textit{Kepler} have done an outstanding job and provided the community with a wealth of data that revolutionised the fields of extra-solar planets and asteroseismology. Data from these satellites revealed, among others, the possibility to distinguish between red-giant branch stars and red-clump stars, to determine differential rotation between the core and the outer layers of the stars, as well as the location of sharp internal structure changes in stars other than the Sun. Nevertheless, there exist still open questions such as, what are the ages of stars, or how do the large variety of planetary systems form. To answer these questions dedicated efforts for several new instruments are underway:
\begin{itemize}
\item BRITE-Constellation (BRIght Target Explorer) is a network of nano-satellites {with different colour filters} to investigate the properties of the brightest stars in the sky \citep{kaiser2012}. The Austrian BRITE-satellites have been launched successfully on February 25, 2013.
\item On April 5, 2013, NASA announced the selection of the Transiting Exoplanet Survey Satellite (TESS). TESS will perform an all-sky survey to discover transiting exoplanets ranging from Earth-sized to gas giants in orbit around the nearest and brightest stars in the sky. Its goal is to identify terrestrial planets in the habitable zones of nearby stars. This mission will also be very useful to obtain accurate asteroseismology for the nearest and brightest stars.
\item PLATO (PLAnetary Transits and Oscillations of stars) is a candidate mission for ESA. The primary goal of PLATO is to open a new way in exoplanetary science, by providing a full statistical analysis and characterisation of exoplanetary systems around stars that are bright and nearby enough to allow for simultaneous detailed studies of their host stars \citep{rauer2012}.
\item From the ground a network of telescopes is under construction. SONG (Stellar Observations Network Group) is a Danish-led project to construct a global network of 1m telescopes. These telescopes have a high-resolution spectrograph dedicated to asteroseismic measurements. The first telescope has been built at Observatorio del Teide, Tenerife, and will enter the operation mode in the summer of 2013 \citep{uytterhoeven2012}. Another telescope of the SONG network is currently under development in China \citep{deng2013}.

Spectroscopic observations of solar-like oscillations have the advantage that the granulation signal is less prominent and hence the oscillations are more prominent. Furthermore, it is more likely to observe $\ell=3$ modes using spectroscopic observations providing additional constraints for the determination of the stellar structures.
\end{itemize}
 
All these efforts are complementary to CoRoT and \textit{Kepler} in both observing strategy and in the target stars. BRITE, TESS, and PLATO will aim at much brighter stars than CoRoT and \textit{Kepler} which will allow for complementary ground-based efforts to determine effective temperatures and metallicities, which are important observables for stellar modeling. Furthermore, these missions will observe different parts of the sky allowing for stellar-population studies of  different fields. TESS will even perform an all-sky survey albeit at the cost of shorter datasets, i.e., lower frequency resolution. PLATO will observe two different large fields ($\sim$ 500 square degrees) for about two to three years each before going to a step and stare mode. This will have the advantage of long datasets for tens of thousands of bright stars, which are currently lacking. Additionally, the synergy with Gaia \citep{torra2013} will allow for accurate distance and luminosity determinations. Finally, BRITE has different colour filters which will be of prime importance for the mode identification of oscillations driven by the heat-engine mechanism.

With the efforts that are currently underway future high-quality asteroseismic observations are secured. These efforts are necessary to fully understand the internal structures of stars including their ages and planet formation scenarios.\newline
\newline
Finally, the CoRoT and \textit{Kepler} legacy is a truly incredible amount of very high quality data. These data are of unprecedented quality in terms of signal-to noise ratio, continuity and total length of the timeseries as well as the vast number and range of different stars. The analysis and interpretation of these data will continue to reveal important results. These add to the wealth of results already obtained and (partly) described in this review, indicating the significant impact of asteroseismic results from CoRoT and \textit{Kepler} on our understanding of stellar structure and evolution. 

\section{Acknowledgements}
SH would like to thank Anwesh Mazumdar and Maarten Mooij for useful discussions and for providing detailed feedback on earlier drafts. {SH also wants to thank the referee R.A. Garc\'ia for critical comments that improved the paper significantly.} SH acknowledges financial support from the Netherlands Organisation for Scientific Research (NWO).

\bibliographystyle{elsarticle-harv}
\bibliography{reviewASR.bib}

%% Authors are advised to submit their bibtex database files. They are
%% requested to list a bibtex style file in the manuscript if they do
%% not want to use elsarticle-harv.bst.

%% References without bibTeX database:

% \begin{thebibliography}{00}

%% \bibitem must have one of the following forms:
%%   \bibitem[Jones et al.(1990)]{key}...
%%   \bibitem[Jones et al.(1990)Jones, Baker, and Williams]{key}...
%%   \bibitem[Jones et al., 1990]{key}...
%%   \bibitem[\protect\citeauthoryear{Jones, Baker, and Williams}{Jones
%%       et al.}{1990}]{key}...
%%   \bibitem[\protect\citeauthoryear{Jones et al.}{1990}]{key}...
%%   \bibitem[\protect\astroncite{Jones et al.}{1990}]{key}...
%%   \bibitem[\protect\citename{Jones et al., }1990]{key}...
%%   \harvarditem[Jones et al.]{Jones, Baker, and Williams}{1990}{key}...
%%

% \bibitem[ ()]{}

% \end{thebibliography}

\end{document}